\documentclass[a4paper,fleqn,usenatbib]{mnras}

\usepackage{newtxtext,newtxmath}

\usepackage[T1]{fontenc}
\usepackage{ae,aecompl}

\usepackage{graphicx}	
\usepackage{amsmath}	
\usepackage{amssymb}	

\title[New mechanism of long-term period variations]{A new mechanism of long-term period variations for W UMa-type contact binaries}

\author[L. Liu et al.]
{L. Liu$^{1,2,3,4}$\thanks{E-mail: LiuL@ynao.ac.cn}, S.-B. Qian$^{1,2,3,4}$, X. Xiong$^{1,2,3,4}$
\\
$^{1}$Yunnan Observatories, Chinese Academy of Sciences, 396 Yangfangwang, Guandu District, Kunming, 650216, P. R. China\\
$^{2}$Key Laboratory for the Structure and Evolution of Celestial Objects, Chinese Academy of Sciences,\\ 396 Yangfangwang, Guandu District, Kunming, 650216, P. R. China\\
$^{3}$Center for Astronomical Mega-Science, Chinese Academy of Sciences, 20A Datun Road, Chaoyang District, Beijing, 100012, P. R. China\\
$^{4}$University of Chinese Academy of Sciences, Yuquan Road 19\#, Sijingshang Block, 100049 Beijing, China\\}

\date{Accepted XXX. Received YYY; in original form ZZZ}

\pubyear{2017}

\begin{document}
\label{firstpage}
\pagerange{\pageref{firstpage}--\pageref{lastpage}}
\maketitle

\begin{abstract}
W UMa-type contact binaries belong to close binary systems whose components exactly overflow their Roche lobes and share a common convective envelope (CCE). In the last twenty years, the long-term variations of their orbital periods have been thought to depend on several mechanisms. Now, we suggest a new mechanism: CCE-dominated mechanism. The CCE-dominated mechanism is found based on our numerical result, especially at high mass ratios, that the orbital periods ($P$) of contact binaries change very much with their fill-out factors ($f$). Because $f$ is taken as a measurement of the thickness of CCE, the physical cause for the variation of $P$ is a mass transfer between CCE and components. Further, an $f$-dominated simplification model for this mechanism is introduced. According to it, $P$ may change in a long-term oscillation way with a similar time scale of the thermal modulation, meanwhile $q$ is decreasing slowly till the two components merge. It could be also applied to explain the presence of extremely short period, high mass ratio and deep contact binaries. Moreover, the CCE-dominated mechanism should always work due to mass transfer and mass loss both occurring via CCE. Therefor, the effect of CCE on the variations of orbital periods may have been underestimated before.
\end{abstract}



\section{Introduction}
A contact binary is a close binary system whose components are exactly over their Roche lobes \citep{Kopal1959} and share a common convective or radiative envelope. The late-type contact binary, so-called W UMa-type contact binary, possesses a common convective envelope \citep[CCE, e.g.][]{Lucy1968}. Generally, contact binaries are thought to be formed from the moderate close binaries \citep[e.g.][]{Qianetal2017} via the nuclear evolution, the angular momentum loss \citep[AML, e.g.][]{Huang1967,vantVeer1979,Rahunen1981,Vilhu1982,GuinanBradstreet1988,Hilditchetal1988,vantVeerMaceroni1989,Yildiz2014} or the Kozai effect \citep{Kozai1962} caused by third bodies \citep[e.g.][]{Zascheetal2009,Conroyetal2014}, and finally to be merged into a single star like V1309 Sco \citep{Tylendaetal2011, Zhuetal2016,Pietrukowiczetal2017}. How long it will take from the formation to the merging, however, is still an open issue \citep[from less than 1 Gyr to 8.89 Gyr, e.g.][]{Biliretal2005, Lietal2007, Yildiz2014}.

The mass transfer/loss is the key process that drives the evolution of the close binary systems, as well as the contact binaries. These processes would change the mass distributions and the chemical compositions of binary systems. The mass redistributions will lead to variations of the orbital periods, while the chemical composition redistributions will change their spectra. Rates of these redistributions would determine how long a contact binary system will take from its formation to merger. Here, we focus on variations of orbital periods for W UMa-type contact binaries and on their mechanisms, which would be very important to understand the evolution of this type of binaries.

The increasing or decreasing variations of the orbital periods for W UMa-type contact binaries were observed with an equiprobability \citep[e.g.][]{Kreiner1977,LiaoQian2010, Lohretal2013}, by using the $(O-C)$ method, where ``$O$" is the observed times of minima, while ``$C$" is the calculated times of minima from a theoretical ephemeris. According to the causes, such variations can be classified into two types: apparent variations or intrinsic variations. The apparent variations are caused by the orbital motion, such as light travel time effect (LTTE), apsidal motion, and so on. On the contrary, the intrinsic variations are caused by the changes of components in geometry or in structure. Furthermore, according to the time scales, such variations can be classified into long-term variations and short-term variations.

The long-term variations usually happen on a thermal time scale, while the short-term ones usually happen on a time scale of decades or less. There are many arguments about the short-term variations because the currently observational accuracy is not high enough to test which mechanism is dominant (LTTE, e.g. \citealt{Frieboes-CondeHerczeg1973,Chambliss1992,BorkovitsHegedus1996}, magnetic activities, i.e. Applegate mechanism, e.g. \citealt{Applegate1992,Lanzaetal1998,LanzaRodono1999}, circulation within CCE, e.g. \citealt{Robertson1980,Eaton1983,RucinskiVilhu1983,Rucinski1985,Kahler1989}, mass motions within the components, e.g. \citealt{Webbink1976,Vilhu1981}).

In contrast to the short-term variations, the observed long-term variations are argued less. It is widely accepted that they can be caused by several possible mechanisms as follows:

(1) Mass loss from the binary system, which eternally makes the period increasing \citep[e.g.][]{Sawadawetal1984,GehrzSmith2002,AndradeDocobo2003,Nanourisetal2011},

(2) Mass transfer between the components, which makes the period increasing or decreasing \citep[e.g.][]{Pribullaetal2000,Tranetal2013}, and

(3) Angular momentum loss (AML) from the binary system via a magnetic stellar wind, which permanently makes the period decreasing \citep[e.g.][]{Moss1972,Rucinski1982}.

\noindent If the system possess a strong magnetic field, the mechanisms (1) and (3) would work together, which makes the period decreasing. \cite{ToutHall1991} yielded a formula for these mechanisms.

Generally, during the steady contact phase, the rate of mass loss or of AML is too small to cause such long-term variations \citep{WargelinDrake2002,Woodetal2002, Stepien2006}. The long-term variations therefor are mainly caused by mass transfer. The sign of the transfer rate depends on the direction of the matter flow. The processes of mass transfer determine the evolutionary fates of close binaries. Imaging that a star in the binary system could obtain new fuels continuously from its companion, its evolution must be different from that of a single star with the same mass. In many kinds of close binary systems, the direction of the mass transfer is certain, the rate of this process is the remaining key parameter for dominating their types and fates. In a W UMa-type contact binary, however, the direction must be considered very carefully, because this type binary has a long-lived CCE via which the matter is able to exchange freely from one component to the other \citep[e.g. the TRO thoery,][]{Lucy1976, Flannery1976,RobertsonEggleton1977}. This free mass transfer presents variations of the orbital periods increasing or decreasing in long-term observations.

Several works were attempted to investigate the behavior of the long-term variations \citep[e.g.][]{Qian2001a, Qian2001b,Qian2003}. From these works, it was suggested that the periods of contact binaries change around a critical value of mass ratio (i.e. $q=0.4$), which resulted from the alternate dominance of TRO and the varied rates of AML. These works supported the conjecture \citep[][]{Vilhu1981} that the depth of contact (equivalently the thickness of CCE) affects the intensity of the surface magnetic field and consequently influences the magnetic braking which determines the rate of AML. A thick CCE would restrain the surface magnetic field resulting in a small rate of AML, so that TRO dominated this process and the orbital period increased. Otherwise, the orbital period decreased under AML dominant.

W UMa-type contact binaries possess long-lived steady CCE, which distinguishes them from all other close binaries. However, the long-term influences of CCE on the orbital period have been neglected for a long time. Since the concept of CCE had been put forward by \cite{Lucy1968}, and especially since W UMa-type contact binaries had been divided into A- and W-type by \cite{Binnendijk1970} according to the morphology of their light curves, many follow-up research efforts had only focused on the details of energy transfer within CCE, from thermal equilibrium models to thermal non-equilibrium models, which attempted to resolve the so-called Binnendijk's paradox (the less massive component is the hotter one). \cite{Whelan1972} investigated the situation of energy transfer in the strongly superadiabatic outer part of CCE. This energy transfer ultimately makes it possible that the less massive component is hotter than the more massive one. \cite{BiermannThomas1972, BiermannThomas1973} thought that the transferred energy is carried by the fluid circulation, which could be driven by the condition of considerably different entropy between the two stellar envelopes. \cite{Shuetal1976,Shuetal1979} suggested that CCE can achieve thermal equilibrium but to be unequal specific entropy. Thus, the temperature of CCE is inversive. They called this as the contact discontinuity. \cite{Webbink1977} insisted that large-scale circulation between components which absorbs or releases energy in CCE can be maintained. \cite{Robertson1980} presented a model that energy can be transferred by steady circulation within CCE. \cite{SmithSmith1981} proved that the inertial and/or Coriolis forces can not be neglected when the circulation within CCE was calculated in a realistic model. \cite{Eaton1983} analyzed the short-wavelength IUE spectra of 14 W UMa-type binary systems and found the observed result being inconsistent with the predictions about CCE with the contact-discontinuity model. \cite{Kahler1989} summarized several models of CCE and concluded that the isentropic condition could not be correct if hydrostatic thermal equilibrium were adopted. He believed that turbulence which is driven by large-scale circulation currents and by convective instability plays a central role in the process of energy transfer within CCE. \cite{Tassoul1992} considered a treatment of the 3D barocline in his hydrodynamic model for energy transfer within CCE. He claimed that his model can be applied to both the late-type and the early-type contact binaries because the model did not include lateral convection along the Roche equipotential surface. \cite{Wang1994} proposed that the interaction between the secondary component and CCE, which leads to A- and W-type W UMa-type contact binaries. \cite{ZhouLeung1997} introduced a 2D approach to simulate the circulation within CCE, neglecting the Coriolis effect. Their simulation showed the morphology of the circulation in the neck of a contact binary system. As they pointed out, however, that it is important to simulate the circulation numerically with the Cariolis force in 3D. Even \cite{Hazlehurst1999} applied the flow topology to contact binaries.

All these works mainly focused on the short-term behaviour of CCE, neglecting the long-term part. For this reason, we will discuss the long-term behaviour of CCE in this paper.

\section{A new mechanism for period variation}
As mentioned above, the long-term oscillation behaviour of the orbital periods of W UMa-type contact binaries were widely observed. This behaviour is prevalently thought to be caused by direct mass transfer between the components. Nevertheless, we find a new mechanism which may also cause a similar long-term oscillation on $P$. This mechanism is based on the fact that $P$ could change with the thickness of CCE, while the thickness of CCE is strongly corrected with the fill-out fact $f$. $P$ and $f$ are dynamic parameters of a contact binary system. To simulate such dynamic parameters, we shall start with the Roche geometry.

\subsection{The effective radii of contact binaries}
The prevailing structure models of contact binaries originate from Roche geometry. Based on the definition, the radii of components of a contact binary are well limited by the corresponding radii of the Roche lobes described by Roche potential. Under the assumption of mass point, the Roche potential under an orthogonal coordinate system can be calculated by
\begin{equation}
{\psi}=\frac{2}{1+q}\cdot\frac{1}{r_1}+\frac{2q}{1+q}\cdot\frac{1}{r_2}+\left(x-\frac{q}{1+q}\right)^2+y^2,
	\label{eq:Roche potential}
\end{equation}
\noindent where ${r_1}^2=x^2+y^2+z^2$, ${r_2}^2=(1-x)^2+y^2+z^2$, $q=M_2/M_1$ ($M_2{\leqslant}M_1$), and where $x$, $y$ and $z$ are normalized coordinates (in semi-major axis, $A$).
The Lagrangian point 1 (${\rm{L}}_1$) is a maximum potential point on the x-axis between $M_1$ and $M_2$, while the Lagrangian point 2 (${\rm{L}}_2$) is another maximum potential point on the x-axis outside the less massive component $M_2$. The equipotential surface which passes the ${\rm{L}}_1$ is called inner Lagrangian equipotential surface, and the one which passes the ${\rm{L}}_2$ is called outer Lagrangian equipotential surface. The corresponding potentials of inner and outer Lagrangian equipotential surface are marked as ${\psi}_{\rm{in}}$ and ${\psi}_{\rm{out}}$, respectively. The actual equipotential surface of a contact binary is between the inner and outer Lagrangian equipotential surfaces and can be described by the fill-out factor which is defined by
\begin{equation}
f=\frac{{\psi}-{\psi}_{\rm{in}}}{{\psi}_{\rm{out}}-{\psi}_{\rm{in}}}.
	\label{eq:fill-out factor}
\end{equation}
\noindent Thus, we have ${\psi}={\psi}(x,y,z,q,f)$. For a given $q$ and $f$, the radii of Roche lobes, relating with the coordinates $x$, $y$ and $z$, can be calculated. Firstly, we get the coordinates of ${\rm{L}}_1$ [($x_{1\rm{L}}$, 0, 0), $0<x_{1\rm{L}}<1$] and ${\rm{L}}_2$ [($x_{2\rm{L}}$, 0, 0), $x_{2\rm{L}}>1$] under a given $q$, respectively. Then, substituting the values of $q$, $x_{1\rm{L}}$ and $x_{2\rm{L}}$ in Equation~\ref{eq:Roche potential}, we obtain the values of ${\psi}_{\rm{in}}$ and ${\psi}_{\rm{out}}$, respectively. Further, we obtain the ${\psi}$ with Equation~\ref{eq:fill-out factor} when $f$ is given. Secondly, we calculate the three normalized characteristic radii, so-called pole, side and back radius, of the geometry-determined Roche lobes. The radii mentioned above can be calculated as follows, $r_{1\rm{pole}}={\mid}z_1{\mid}$, ${\psi}(0,0,z_1,q,f)={\psi}$, $r_{2\rm{pole}}={\mid}z_2{\mid}$, ${\psi}(1,0,z_2,q,f)={\psi}$, $r_{1\rm{side}}={\mid}y_1{\mid}$, ${\psi}(0,y_1,0,q,f)={\psi}$, $r_{2\rm{side}}={\mid}y_2{\mid}$, ${\psi}(1,y_2,0,q,f)={\psi}$, $r_{1\rm{back}}={\mid}x_1{\mid}$, ${\psi}(x_1,0,0,q,f)={\psi}$, $x_1<0$, and $r_{2\rm{back}}=x_2$, ${\psi}(x_2,0,0,q,f)={\psi}$, $1<x_2\leqslant x_{2\rm{L}}$. The universal radius of a Roche lobe can be calculated by $r_i=\sqrt[3]{r_{i\rm{pole}}{\cdot}r_{i\rm{side}}{\cdot}r_{i\rm{back}}}$. At last, we obtain the radii of Roche lobes, namely the normalized radii of each component with certain $q$ and $f$. Thirdly, after setting $q$ from 0.001 to 1.000 with a step of 0.001, and setting $f$ from 0.0\,\% to 100.0\,\% with a step of 10.0\,\%, we obtain a set of values of $q$, $f$, $r_1=r_1(q,f)$ and $r_2=r_2(q,f)$. The real radius of each component should be the value of the normalized radius times the semi-major axis ($A$), namely, $R_1=A{\cdot}r_1(q,f)$, and $R_2=A{\cdot}r_2(q,f)$.

\subsection{Calculation of the dynamic parameters of contact binaries}
\begin{figure}
	\includegraphics[width=\columnwidth]{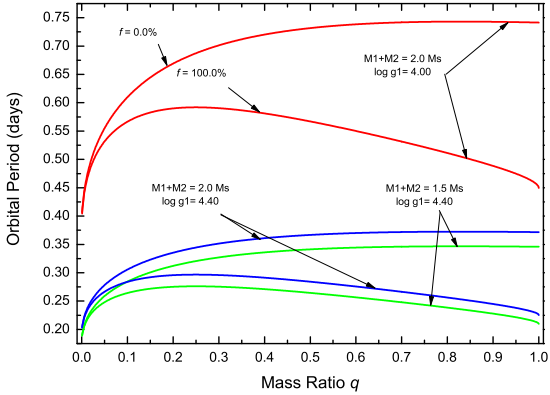}
\caption{The contact binary dynamic parameter map. The solid lines denote the relationship between $q$, $P$ and $f$ with constants $M$ and log\,$g_1$. The same color lines constitute a couple, which denotes a field from $f=0$\% to $f=100$\%.} \label{fig:q-p-f-M-logg}
\end{figure}

\begin{figure}
	\includegraphics[width=\columnwidth]{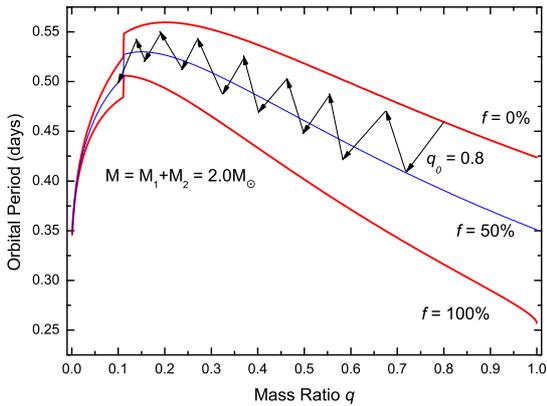}
\caption{The contact binary dynamic evolution map. The solid lines are the same as the Fig~\ref{fig:q-p-f-M-logg}. In this figure, the log\,$g_1$ is satisfied with the relationship of a main sequence stars (Equations~\ref{eq:R-M relation}). A contact binary system should be observed in the section limited by the red lines in its whole contact phase if the total mass is conservative. The black line with arrows, which is yielded by the simplification model in section 2.4, is a suppositional evolutionary track of the contact binary ($q_0=0.8$) with the total mass of two solar masses. The jumps of the red and blue lines are caused by the discontinuity of Equations~\ref{eq:R-M relation}.} \label{fig:q-p-f-M}
\end{figure}

To obtain the dynamic parameters of a contact binary, we assume that the total mass of the system is $M$ ($M=M_1+M_2$), and that the surface gravity acceleration of the more massive component is $g_1$ (log\,$g_1={\rm{log}}\,[M_1/R_1^2]+4.43$, in solar units). Hence, the radius of the massive component, $R_1$, is computed by
\begin{equation}
R_1=\sqrt{M_1 \times 10^{4.43-{\rm{log}}\,g_1}},
   \label{eq:R1}
\end{equation}
\noindent where $M=M_1+M_2,~M_1=M/(1+q)$. Then, by the definition of $r_1(q,f)$ and $r_2(q,f)$, we obtain
\begin{equation}
A=R_1/r_1(q,f),~R_2=A{\cdot}r_2(q,f).
   \label{eq:A}
\end{equation}
\noindent Similarly, the surface gravity acceleration of the less massive component can be calculated by
\begin{equation}
	{\rm{log}}\,g_2= {\rm{log}}\,(M_2/R_2)^2+4.43,
    \label{eq:logg2}
\end{equation}
\noindent where $M_2=M{\cdot}q/(1+q)$. According to the Kepler's third law, we have
\begin{equation}
P=0.1159A^{\frac{3}{2}}M^{-\frac{1}{2}}=243.4{\times}10^{-0.75{\rm{log}}\,g_1}M^{\frac{1}{4}}(1+q)^{-\frac{3}{4}}[r_1(q,f)]^{-\frac{3}{2}},
    \label{eq:P}
\end{equation}
\noindent where $P$ is in days, $A$ is in solar radius and $M$ is in solar mass. Finally, we obtain the relationship between the orbital period $P$ and the mass ratio $q$, with certain $f$, $M$ and log\,$g_1$. We call this relationship as the contact binary dynamic parameter map (See Fig~\ref{fig:q-p-f-M-logg}). It should be noted that the solid lines in Fig~\ref{fig:q-p-f-M-logg} are not real evolutionary tracks for contact binaries, because log\,$g_1$ of the more massive component should change with the changing mass either by nuclear evolution or by mass transfer. To obtain more reliable tracks, we assume that log\,$g_1$ obeys the corresponding empirical relationship of the main sequence according to the observational facts that the more massive components of contact binary systems are almost located on the zero age main sequence \citep[ZAMS, e.g.][]{YakutEggleton2005, YildizDogan2013}. Here, we use one set of formulae of the M-R relation for MS \citep[e.g.][]{GimenezZamorano1985}
\begin{equation}
\begin{cases}
R_1&=10^{0.053+0.997{\rm log}M_1}, M_1<1.8\,{\rm M_{\odot}},\\
R_1&=10^{0.153+0.556{\rm log}M_1}, M_1 \geqslant1.8\,{\rm M_{\odot}},
\end{cases}
    \label{eq:R-M relation}
\end{equation}
\noindent to replace Equation~\ref{eq:R1}, yielding evolutionary tracks of a contact binary system ($M=2.0\,{\rm M_{\odot}}$) as shown in Fig~\ref{fig:q-p-f-M}. It is emphasized that which M-R relation used here is NOT important. We just want to use an M-R relation of the main sequence stars to illustrate the evolutionary track of a contact binary system whose log\,$g_1$ is not a constant.

\begin{figure}
	\includegraphics[width=\columnwidth]{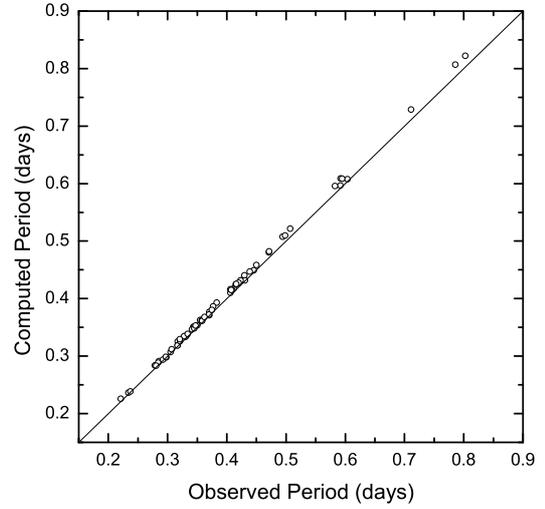}
\caption{Observed versus Computed orbital periods for contact binary samples of \citet{YakutEggleton2005}.} \label{fig:OCP}
\end{figure}

To check our computations, the samples of contact binaries collected by \citet{YakutEggleton2005} are used. These samples are reliable, as \cite{YakutEggleton2005} said, because of both spectroscopic and photometric analysis being available. The corresponding results of the examination are listed in Table~\ref{tab:sample}. The first to the seventh columns are totally the same as the table\,1 of that paper \citep{YakutEggleton2005}, while the eighth to the eleventh columns are calculated from the corresponding masses and radii. The last four columns are the calculated values of orbital periods (c\_Per, by Equation~\ref{eq:P}) and of surface gravity accelerations for the less massive component (c\_log\,$g_2$, by Equation~\ref{eq:logg2}), and their relative deviations from the observational values (d\_c\_Per and d\_c\_log\,$g_2$). Relative deviations are calculated with the formula $\rm{\mid}Comp-Obs{\mid}/Obs$. The relative deviations of the orbital periods are less than 3\%. When the calculated parameters are replaced by the surface gravity accelerations, the relative deviations are sufficiently smaller than 0.7\%. Fig~\ref{fig:OCP} is the diagram of observed versus computed orbital periods for these samples. Most of samples are located on the diagonal. Hence, the results of our computations are reliable and correct.

\subsection{Variation of orbital period}
In the $q-P-f$ diagram (Fig~\ref{fig:q-p-f-M}), the upper line is the boundary of the fill-out factor $f=0$, while the lower one with the same colour is the boundary $f=1$. A broad evolutionary space is formed between these two lines. In this space, $P$ permanently decreases as $f$ increases. When $q$ is quite small, which implies that the contact binary system is going to merge, $P$ decreases very fast. As shown in Fig~\ref{fig:q-p-f-M}, $P$ changes from 0.45 to 0.56 days when $q$ changes from 1.0 to 0.2 at $f = 0$, while $P$ changes from 0.45 to 0.26 days when $f$ changes from 0 to 1 at $q = 1$. It reveals that $P$ can vary in two ways: mainly with $q$ or mainly with $f$. The first way has been studied very much and been widely accepted. The second way, which is comparable to the first one, also has its physical meaning. Note that $f$ is an indicator for the thickness of CCE. When matter enters or leaves CCE, the thickness of CCE varied, and therefor $f$ and $P$. Hence, the period variation in the second way are caused by mass transfer between components and CCE.

This variation can be understood as the material redistribution in the Roche lobes. The barycentres change due to the mass redistribution, so that the orbital period changes with the varied separation. It can also be understood as transfer of angular momentum (AM). The rotational inertia of each component should become larger when matter enters into CCE. This process should cause a larger spin AM so that the orbital AM should decrease. As a result, the orbit is shrinking while $P$ is decreasing. Anyway, if the thickness of CCE changed, the period would indeed change correspondingly. Over a long time, the accumulative variations should be obvious enough to be observed. It may be a new mechanism for the long-term variation of periods of W UMa-type contact binaries.

\subsection{An $f$-dominated simplification model }
According to the new mechanism, we introduce a simplification model in this section, in which $f$ is dominant in long-term period variations.

Mark the initial mass of the donor and the acceptor as $M_{\rm don0}$ and $M_{\rm acc0}$, respectively, so the initial mass ratio is $q_0=M_{\rm don0}/M_{\rm acc0}$. The mechanism is starting from a system with its $q_0$. From there, it has a two-phase process. In the first phase the direction of mass transfer is from the donor to CCE, $f$ being increasing. If the initial fill-out factor is $f_0$ with an increment of ${\Delta}f_{\rm{ph1}}$, we have, at the end of phase one,
\begin{equation}
	f_{\rm{ph1}}=f_0+{\Delta}f_{\rm{ph1}}.
    \label{eq:fph1}
\end{equation}
\noindent Note that the donor and the acceptor will share the increasing matter in CCE. It means that a part of the increasing matter in CCE goes back to the donor, and the other part goes to the acceptor, because the mass of each component should include the mass of their respective CCE. Hence, it should assume that the lost mass of the donor is ${\Delta}M_{\rm{dph1}}$, and the increased mass of the acceptor is ${\Delta}M_{\rm{aph1}}$, respectively. Consequently,
\begin{equation}
	q_{\rm{ph1}}=\frac{M_{\rm don0}-{\Delta}M_{\rm{dph1}}}{M_{\rm acc0}+{\Delta}M_{\rm{aph1}}}.
    \label{eq:qph1}
\end{equation}
In this phase $f$ is increasing and $q$ is decreasing. As shown in the paper with the numerical simulation, $P$ is decreasing with the increasing $f$.

In the second phase the direction of mass transfer is from CCE to the acceptor. For the same reason as the first phase, it could assume that, in this phase, the decreased mass of the donor is ${\Delta}M_{\rm{dph2}}$, and the gained mass of the acceptor is ${\Delta}M_{\rm{aph2}}$, respectively. Therefor,
\begin{equation}
	f_{\rm{ph2}}=f_{\rm{ph1}}-{\Delta}f_{\rm{ph2}},
    \label{eq:fph2}
\end{equation}
\begin{equation}
	q_{\rm{ph2}}=\frac{M_{\rm don0}-{\Delta}M_{\rm{dph1}}-{\Delta}M_{\rm{dph2}}}{M_{\rm acc0}+{\Delta}M_{\rm{aph1}}+{\Delta}M_{\rm{aph2}}}.
    \label{eq:qph2}
\end{equation}
\noindent Both $f$ and $q$ are decreasing in the second phase, while $P$ is increasing.

Because the mass of the fully filled CCE is only a small fractional of the total mass, not too much matter will be transferred from one component to the other in one cycle of this two-phase process. This implies that it will take several cycles to yield an extreme low mass ratio contact binary system which is a candidate of the merger because of dynamic instability \citep[e.g.][]{Hut1980,Rasio1995}. As a result, this two-phase process will be lasting, where $q$ is constantly decreasing, till it is low enough to be satisfied the merging condition while $f$ is oscillating from low to high. Since $P$ is strongly correlated with $f$, the mechanism explaining the variation of $f$, can also explain the variation of $P$.

\subsection{Illustration of the model}
The situations of a real contact binary system are very complicated. To fully understand the evolution of contact binaries, many detailed investigations are needed, which are beyond the scope of this work. Hence, we just show an instance to illustrate our model. Based on some currently observational facts, we assume a conservative contact binary system as $q_0=0.8$, $M=M_{\rm don0}+M_{\rm acc0}=2\,\rm M_\odot$.

In the first phase, assuming $f$ varies from 0 to 0.5, and thus ${\Delta}f_{\rm{ph1}}=0.5$, and the corresponding ${\Delta}P_{\rm{ph1}}=0.065$\,days according to Fig~\ref{fig:q-p-f-M}. A typical observed value of $dP/dt$ is $10^{-7}$\,d/yr \citep[e.g.][]{Lohretal2013}. Hence, the duration time of the phase one is $6.5\times10^5$\,yr, and then $df/dt=7.69\times10^{-7}$\,yr$^{-1}$, which is too small to be detected in a few decades, even hundreds, unless a zoom method like the $(O-C)$ be found. Assuming the mass fraction of CCE is 10\%, and assuming the donor and the acceptor share CCE with a ratio which is similar to $q$, so ${\Delta}M_{\rm{dph1}}=0.0445\,\rm M_\odot$, ${\Delta}M_{\rm{aph1}}=0.0555\,\rm M_\odot$, and $q_{\rm{ph1}}=0.725$, based on Equation~\ref{eq:qph1}.

In the second phase, assuming $f$ varies from 0.5 to 0.1, the corresponding ${\Delta}f_{\rm{ph2}}=0.4$ and ${\Delta}P_{\rm{ph2}}=0.062$\,days. By using the same values and assumptions as above, the duration time of the phase two is $6.2\times10^5$\,yr, ${\Delta}M_{\rm{dph2}}=0.0336\,\rm M_\odot$, ${\Delta}M_{\rm{aph2}}=0.0464\,\rm M_\odot$, and $q_{\rm{ph2}}=0.670$ , based on Equation~\ref{eq:qph2}.

The first two-fragment of the black-arrow lines in Fig~\ref{fig:q-p-f-M} are these corresponding results. Similarly, the remaining black-arrow lines are obtained. Moreover, the total duration time which the contact binary will be taken to evolve along the black-arrow lines can be calculated. It approximates the life-time of the contact binary system.

\subsection{Some physical details of the model}
\subsubsection{The donor and the acceptor}
Generally, the matter flow should be always from the high to the low density section if the other conditions are similar. Because the components of a contact binary exactly filled their Roche lobes, the ratio of the mean density depends on $q$. According to the Roche model, the mean density of the less massive component ($\overline{\rho}_2$) is always greater than that of the more massive one ($\overline{\rho}_1$). On the other hand, the geometrical size of the primary component (more massive) is also always larger than that of the secondary (less massive). Bigger geometrical size means sparse Roche potential surfaces, therefor a smaller density gradient. Thus, the density gradient of the primary component should be smaller than that of the secondary one, and hence the matter falling into the core of the primary one could be more easier than it falling into the core of the secondary one. As a result, the secondary component should be the donor, while the primary one should be the acceptor, all the time.

\subsubsection{A possible physical interpretation of the two-phase process}
Actually, the mass transfer between the components usually occurs via CCE. In most situations, the overflowing matter from the donor enters CCE rather than enters the other component directly, because of the rotation and the geometric construction. A simple fact that the direct channel of mass transfer between the components is a narrow neck through the Lagrangian point 1 (${\rm{L}}_1$), while the channel between CCE and components is almost the whole surface. Mass transfer in the latter way should be more easy. Probably, this overflowing matter can be accumulated in CCE till a critical condition (the first phase), and then it is released to the acceptor (the second phase). The critical condition may be related to the mean density of CCE ($\overline{\rho}_{\rm CCE}$). Namely, the criteria may be that if $\overline{\rho}_{\rm CCE} < \overline{\rho}_1 < \overline{\rho}_2$, the first phase is dominant, while if $\overline{\rho}_1 < \overline{\rho}_{\rm CCE} < \overline{\rho}_2$, the second phase is dominant. The time scales of these two phases may be similar to the corresponding thermal time scales of each component. Thus, the contact binary would evolve along the black-arrow lines in Fig~\ref{fig:q-p-f-M}.

\subsection{An application for explaining the peculiar population}
\begin{figure}
	\includegraphics[width=\columnwidth]{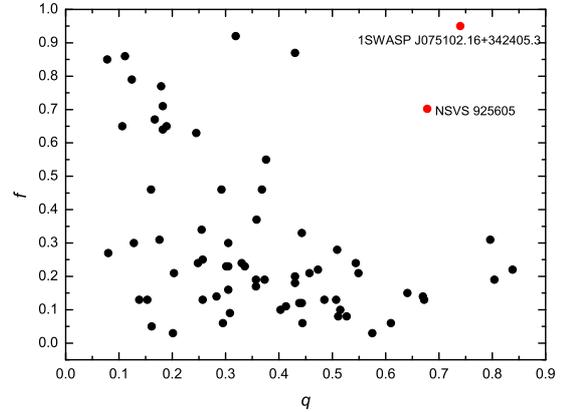}
\caption{The relationship between $q$ and $f$ for contact binary samples of \citet{YakutEggleton2005}. The two red solid circles denote the two extremely short period, deep fill-out factor and high mass ratio contact binaries.} \label{fig:qf}
\end{figure}

In Fig~\ref{fig:qf}, it shows the relationship of $q$ and $f$, with the samples from \citet{YakutEggleton2005}. The contact binaries, whose mass ratios are higher than 0.5 meanwhile whose fill-out factors are greater than 0.3, are very rare in this figure, only two targets, which are extremely short period, high mass ratio and deep contact binaries \citep[e.g.][]{DimitrovKjurkchieva2015, Jiangetal2015}. This distribution may be explained as that contact binaries evolves from high to low mass ratios, with their fill-out factors being increasing; some contact binaries are formed at relatively lower mass ratios so that there are some low mass ratio contact binaries with shallow fill-out factors. As a supplement, with our mechanism, it could be explained the presence of the ``stragglers", which are marked as the red solid circles in Fig~\ref{fig:qf}. According to the mechanism, $P$ becomes very short when $f$ is large at a high value of $q$. This is highly consist with the properties of extremely short period, deep fill-out factor and high mass ratio.

\section{Conclusions}
Through the numerical method, we found that the orbital period of a contact binary system should vary very much when matter moves into or moves off CCE. We proposed an $f$-dominated simplification model to describe this new mechanism, yielding a long-term oscillation variation of $P$. It has also been applied to make an explanation for the presence of the extremely short period, high mass ratio and deep contact binaries. As discussed, the CCE-dominated mechanism ought not to be ignored in most situations. Hence, it is suggested that CCE, which may be the key in the evolution of W UMa-type contact binaries, would be investigated sufficiently in future.

\section*{Acknowledgements}
We are grateful to the anonymous referee who has offered very useful suggestions to improve the paper. This work is partly supported by the Yunnan Natural Science Foundation (2016FB004), the young academic and technology leaders project of Yunnan Province (No. 2015HB098), Chinese Natural Science Foundation (Nos. 11773066, 11403095 and 11325315), the Key Research Programme of the Chinese Academy of Sciences (grant No. KGZD-EW-603).








\clearpage
\begin{table}

\begin{tiny}
\caption{Some contact binary samples for checking our computations. All samples come from the paper of \citet{YakutEggleton2005}.}
	\label{tab:sample}
	\centering
\begin{tabular}{lcccccccccccccc}
\hline
Name            &Per	    & $M_1$ & $M_2$ & $R_1$ & $R_2$ &  $f$    &log\,$g_1$&log\,$g_2$&$M$&$q$    &c\_Per&d\_c\_Per&c\_log\,$g_2$  &d\_c\_log\,$g_2$ \\
                &(days)	    & $(\rm M_\odot)$ & ($\rm M_\odot)$ & ($\rm R_\odot)$ & $(\rm R_\odot)$ &      &  &   &$(\rm M_\odot$)&($M_2/M_1$)&(days) & (\%) &  &(\%) \\
\hline
QX And          &0.4118 	&1.18 	&0.24 	&1.40 	&0.70 	&0.21     &4.210 	&4.120 	&1.42 	&0.203 	&0.41751 	&1.387 	&4.135 	   &0.364 \\
AB And          &0.3319 	&1.01 	&0.49 	&1.05 	&0.77 	&0.13     &4.392 	&4.347 	&1.50 	&0.485 	&0.33564 	&1.127 	&4.363 	   &0.363 \\
GZ And          &0.3050 	&1.12 	&0.59 	&1.01 	&0.76 	&0.08     &4.471 	&4.439 	&1.71 	&0.527 	&0.30714 	&0.702 	&4.448 	   &0.198 \\
OO Aql          &0.5068 	&1.05 	&0.88 	&1.40 	&1.30 	&0.22     &4.159 	&4.147 	&1.93 	&0.838 	&0.52188 	&2.976 	&4.150 	   &0.082 \\
V417 Aql        &0.3703 	&1.40 	&0.50 	&1.31 	&0.84 	&0.19     &4.342 	&4.280 	&1.90 	&0.357 	&0.37701 	&1.812 	&4.295 	   &0.341 \\
SS Ari          &0.4060 	&1.31 	&0.40 	&1.37 	&0.82 	&0.16     &4.274 	&4.204 	&1.71 	&0.305 	&0.41017 	&1.027 	&4.223 	   &0.442 \\
AH Aur          &0.4941 	&1.68 	&0.28 	&1.89 	&0.91 	&0.67     &4.102 	&3.959 	&1.96 	&0.167 	&0.50769 	&2.750 	&3.963 	   &0.099 \\
V402 Aur        &0.6035 	&1.64 	&0.33 	&1.98 	&0.96 	&0.03     &4.052 	&3.984 	&1.97 	&0.201 	&0.60754 	&0.669 	&3.993 	   &0.227 \\
TY Boo          &0.3171 	&0.93 	&0.40 	&1.13 	&0.83 	&0.87     &4.292 	&4.194 	&1.33 	&0.430 	&0.32556 	&2.668 	&4.184 	   &0.236 \\
TZ Boo          &0.2976 	&0.72 	&0.11 	&0.97 	&0.43 	&0.13     &4.314 	&4.204 	&0.83 	&0.153 	&0.29778 	&0.060 	&4.232 	   &0.655 \\
XY Boo          &0.3705 	&0.93 	&0.15 	&1.21 	&0.54 	&0.05     &4.233 	&4.141 	&1.08 	&0.161 	&0.37108 	&0.157 	&4.161 	   &0.476 \\
CK Boo          &0.3551 	&1.42 	&0.15 	&1.48 	&0.59 	&0.65     &4.242 	&4.064 	&1.57 	&0.106 	&0.36265 	&2.126 	&4.080 	   &0.384 \\
EF Boo          &0.4295 	&1.61 	&0.82 	&1.50 	&1.13 	&0.28     &4.285 	&4.238 	&2.43 	&0.509 	&0.44059 	&2.582 	&4.248 	   &0.244 \\
AO Cam          &0.3299 	&1.12 	&0.49 	&1.09 	&0.76 	&0.12     &4.404 	&4.359 	&1.61 	&0.438 	&0.33364 	&1.134 	&4.371 	   &0.285 \\
DN Cam          &0.4983 	&1.85 	&0.82 	&1.76 	&1.25 	&0.33     &4.206 	&4.150 	&2.67 	&0.443 	&0.50981 	&2.310 	&4.159 	   &0.217 \\
TX Cnc          &0.3830 	&0.91 	&0.50 	&1.13 	&0.87 	&0.21     &4.283 	&4.250 	&1.41 	&0.549 	&0.39294 	&2.595 	&4.254 	   &0.096 \\
BH Cas          &0.4059 	&0.74 	&0.35 	&1.11 	&0.80 	&0.22     &4.209 	&4.168 	&1.09 	&0.473 	&0.41619 	&2.535 	&4.173 	   &0.123 \\
V523 Cas        &0.2337 	&0.75 	&0.38 	&0.75 	&0.56 	&0.13     &4.555 	&4.513 	&1.13 	&0.507 	&0.23640 	&1.155 	&4.527 	   &0.301 \\
RR Cen          &0.6060 	&2.09 	&0.45 	&2.24 	&1.07 	&         &4.050 	&4.024 	&2.54 	&0.215 	& 	 	 	  &       &          &      \\
V752 Cen        &0.3700 	&1.30 	&0.40 	&1.27 	&0.75 	&0.09     &4.336 	&4.282 	&1.70 	&0.308 	&0.37250 	&0.676 	&4.291 	   &0.212 \\
V757 Cen        &0.3432 	&0.88 	&0.59 	&1.01 	&0.85 	&0.14     &4.366 	&4.342 	&1.47 	&0.670 	&0.35104 	&2.284 	&4.349 	   &0.161 \\
VW Cep          &0.2783 	&0.93 	&0.40 	&0.93 	&0.64 	&0.18     &4.462 	&4.420 	&1.33 	&0.430 	&0.28392 	&2.019 	&4.424 	   &0.097 \\
TW Cet          &0.3169 	&1.06 	&0.61 	&1.00 	&0.78 	&0.03     &4.455 	&4.431 	&1.67 	&0.575 	&0.31879 	&0.596 	&4.437 	   &0.132 \\
RW Com          &0.2373 	&0.56 	&0.20 	&0.71 	&0.46 	&0.17     &4.476 	&4.406 	&0.76 	&0.357 	&0.23876 	&0.615 	&4.431 	   &0.578 \\
RZ Com          &0.3385 	&1.23 	&0.55 	&1.12 	&0.78 	&         &4.421 	&4.386 	&1.78 	&0.447 	& 	 	 	  &       &          &      \\
CC Com          &0.2210 	&0.79 	&0.43 	&0.75 	&0.58 	&0.24     &4.578 	&4.537 	&1.22 	&0.544 	&0.22598 	&2.253 	&4.547 	   &0.229 \\
eps CrA         &0.5914 	&1.72 	&0.22 	&2.12 	&0.88 	&0.30     &4.013 	&3.883 	&1.94 	&0.128 	&0.59656 	&0.873 	&3.904 	   &0.529 \\
YY CrB          &0.3766 	&1.43 	&0.35 	&1.45 	&0.82 	&0.63     &4.263 	&4.146 	&1.78 	&0.245 	&0.38667 	&2.674 	&4.152 	   &0.134 \\
SX Crv          &0.3166 	&1.25 	&0.10 	&1.31 	&0.44 	&0.27     &4.292 	&4.143 	&1.35 	&0.080 	&0.31841 	&0.572 	&4.157 	   &0.336 \\
DK Cyg          &0.4707 	&1.74 	&0.53 	&1.70 	&1.02 	&0.30     &4.210 	&4.137 	&2.27 	&0.305 	&0.48053 	&2.088 	&4.147 	   &0.240 \\
V401 Cyg        &0.5827 	&1.68 	&0.49 	&1.98 	&1.19 	&0.46     &4.062 	&3.969 	&2.17 	&0.292 	&0.59584 	&2.255 	&3.981 	   &0.300 \\
V1073 Cyg       &0.7859 	&1.60 	&0.51 	&2.51 	&1.64 	&0.92     &3.835 	&3.708 	&2.11 	&0.319 	&0.80684 	&2.664 	&3.693 	   &0.401 \\
V2150 Cyg       &0.5919 	&2.35 	&1.89 	&2.02 	&1.84 	&0.19     &4.190 	&4.177 	&4.24 	&0.804 	&0.60912 	&2.909 	&4.180 	   &0.076 \\
RW Dor          &0.2855 	&0.64 	&0.43 	&0.80 	&0.67 	&0.13     &4.430 	&4.411 	&1.07 	&0.672 	&0.29113 	&1.972 	&4.414 	   &0.061 \\
BV Dra          &0.3501 	&1.04 	&0.43 	&1.11 	&0.75 	&0.11     &4.356 	&4.313 	&1.47 	&0.413 	&0.35380 	&1.057 	&4.321 	   &0.177 \\
BW Dra          &0.2922 	&0.92 	&0.26 	&0.98 	&0.56 	&0.14     &4.411 	&4.349 	&1.18 	&0.283 	&0.29405 	&0.633 	&4.358 	   &0.216 \\
EF Dra          &0.4240 	&1.81 	&0.29 	&1.72 	&0.80 	&0.46     &4.217 	&4.086 	&2.10 	&0.160 	&0.43161 	&1.795 	&4.103 	   &0.411 \\
FU Dra          &0.3067 	&1.17 	&0.29 	&1.13 	&0.62 	&0.24     &4.392 	&4.308 	&1.46 	&0.248 	&0.31197 	&1.718 	&4.324 	   &0.380 \\
YY Eri          &0.3210 	&1.54 	&0.62 	&1.20 	&0.80 	&0.10     &4.459 	&4.416 	&2.16 	&0.403 	&0.32618 	&1.614 	&4.424 	   &0.176 \\
QW Gem          &0.3581 	&1.31 	&0.44 	&1.26 	&0.79 	&0.23     &4.347 	&4.278 	&1.75 	&0.336 	&0.36206 	&1.106 	&4.294 	   &0.369 \\
V728 Her        &0.4713 	&1.65 	&0.30 	&1.81 	&0.92 	&0.71     &4.132 	&3.980 	&1.95 	&0.182 	&0.48220 	&2.313 	&3.992 	   &0.313 \\
V829 Her        &0.3581 	&0.86 	&0.37 	&1.07 	&0.74 	&0.20     &4.306 	&4.260 	&1.23 	&0.430 	&0.36300 	&1.368 	&4.267 	   &0.170 \\
V842 Her        &0.4190 	&1.36 	&0.35 	&1.46 	&0.81 	&0.25     &4.235 	&4.157 	&1.71 	&0.257 	&0.42637 	&1.759 	&4.168 	   &0.262 \\
EZ Hya          &0.4497 	&1.37 	&0.35 	&1.55 	&0.87 	&0.34     &4.186 	&4.095 	&1.72 	&0.255 	&0.45832 	&1.917 	&4.110 	   &0.366 \\
FG Hya          &0.3278 	&1.44 	&0.16 	&1.42 	&0.59 	&0.86     &4.284 	&4.092 	&1.60 	&0.111 	&0.33465 	&2.090 	&4.091 	   &0.035 \\
SW Lac          &0.3207 	&0.98 	&0.78 	&1.03 	&0.94 	&0.31     &4.396 	&4.376 	&1.76 	&0.796 	&0.32966 	&2.794 	&4.382 	   &0.141 \\
XY Leo          &0.2841 	&0.82 	&0.50 	&0.86 	&0.69 	&0.06     &4.475 	&4.451 	&1.32 	&0.610 	&0.28879 	&1.651 	&4.458 	   &0.151 \\
AP Leo          &0.4304 	&1.46 	&0.43 	&1.46 	&0.85 	&0.06     &4.266 	&4.205 	&1.89 	&0.295 	&0.43193 	&0.355 	&4.222 	   &0.413 \\
VZ Lib          &0.3583 	&1.48 	&0.38 	&1.33 	&0.73 	&0.13     &4.353 	&4.283 	&1.86 	&0.257 	&0.36135 	&0.851 	&4.297 	   &0.324 \\
UV Lyn          &0.4150 	&1.36 	&0.50 	&1.45 	&0.96 	&0.46     &4.241 	&4.164 	&1.86 	&0.368 	&0.42475 	&2.349 	&4.172 	   &0.182 \\
TV Mus          &0.4457 	&0.94 	&0.13 	&1.41 	&0.59 	&0.13     &4.105 	&4.002 	&1.07 	&0.138 	&0.44935 	&0.819 	&4.017 	   &0.369 \\
V502 Oph        &0.4534 	&1.38 	&0.48 	&1.45 	&0.89 	&         &4.247 	&4.212 	&1.86 	&0.348 	& 	 	 	  &       &          &      \\
V508 Oph        &0.3448 	&1.01 	&0.52 	&1.07 	&0.80 	&0.10     &4.376 	&4.340 	&1.53 	&0.515 	&0.35001 	&1.511 	&4.351 	   &0.258 \\
V566 Oph        &0.4096 	&1.40 	&0.33 	&1.47 	&0.79 	&         &4.241 	&4.153 	&1.73 	&0.236 	& 	 	 	  &       &          &      \\
V839 Oph        &0.4090 	&1.64 	&0.50 	&1.50 	&0.90 	&0.23     &4.293 	&4.220 	&2.14 	&0.305 	&0.41505 	&1.479 	&4.236 	   &0.368 \\
V2388 Oph       &0.8023 	&1.80 	&0.34 	&2.64 	&1.35 	&0.65     &3.842 	&3.701 	&2.14 	&0.189 	&0.82247 	&2.514 	&3.713 	   &0.329 \\
ER Ori          &0.4234 	&1.53 	&0.98 	&1.40 	&1.15 	&0.15     &4.322 	&4.300 	&2.51 	&0.641 	&0.43170 	&1.960 	&4.303 	   &0.074 \\
U Peg           &0.3748 	&1.15 	&0.38 	&1.25 	&0.78 	&0.24     &4.297 	&4.226 	&1.53 	&0.330 	&0.38054 	&1.531 	&4.243 	   &0.412 \\
BX Peg          &0.2804 	&1.02 	&0.38 	&0.97 	&0.63 	&0.19     &4.465 	&4.411 	&1.40 	&0.373 	&0.28326 	&1.020 	&4.420 	   &0.202 \\
AE Phe          &0.3624 	&1.38 	&0.63 	&1.26 	&0.90 	&0.21     &4.369 	&4.321 	&2.01 	&0.457 	&0.36827 	&1.620 	&4.332 	   &0.258 \\
OU Ser          &0.2968 	&1.02 	&0.18 	&1.09 	&0.52 	&0.31     &4.364 	&4.253 	&1.20 	&0.176 	&0.29874 	&0.654 	&4.271 	   &0.417 \\
Y Sex           &0.4198 	&1.21 	&0.22 	&1.50 	&0.75 	&0.64     &4.161 	&4.022 	&1.43 	&0.182 	&0.42800 	&1.953 	&4.031 	   &0.216 \\
RZ Tau          &0.4157 	&1.70 	&0.64 	&1.58 	&1.07 	&0.55     &4.263 	&4.177 	&2.34 	&0.376 	&0.42546 	&2.348 	&4.187 	   &0.230 \\
EQ Tau          &0.3413 	&1.22 	&0.54 	&1.15 	&0.80 	&0.12     &4.395 	&4.356 	&1.76 	&0.443 	&0.34672 	&1.588 	&4.363 	   &0.156 \\
V781 Tau        &0.3449 	&1.24 	&0.55 	&1.15 	&0.80 	&0.06     &4.402 	&4.364 	&1.79 	&0.444 	&0.34838 	&1.009 	&4.374 	   &0.225 \\
AQ Tuc          &0.5948 	&1.93 	&0.69 	&2.05 	&1.33 	&0.37     &4.092 	&4.021 	&2.62 	&0.358 	&0.60848 	&2.300 	&4.030 	   &0.220 \\
W UMa           &0.3340 	&1.35 	&0.69 	&1.15 	&0.85 	&0.08     &4.439 	&4.410 	&2.04 	&0.511 	&0.33878 	&1.431 	&4.415 	   &0.113 \\
AA UMa          &0.4681 	&1.56 	&0.85 	&1.47 	&1.11 	&         &4.288 	&4.269 	&2.41 	&0.545 	& 	 	 	  &       &          &      \\
AW UMa          &0.4387 	&1.79 	&0.14 	&1.90 	&0.68 	&0.85     &4.125 	&3.911 	&1.93 	&0.078 	&0.44705 	&1.903 	&3.913 	   &0.048 \\
HV UMa          &0.7108 	&2.80 	&0.50 	&2.86 	&1.44 	&0.77     &3.964 	&3.812 	&3.30 	&0.179 	&0.72904 	&2.566 	&3.814 	   &0.046 \\
AH Vir          &0.4075 	&1.36 	&0.41 	&1.41 	&0.84 	&0.23     &4.265 	&4.194 	&1.77 	&0.301 	&0.41500 	&1.840 	&4.207 	   &0.305 \\
GR Vir          &0.3470 	&1.37 	&0.17 	&1.43 	&0.62 	&0.79     &4.256 	&4.076 	&1.54 	&0.124 	&0.35331 	&1.818 	&4.082 	   &0.155 \\
\hline
\end{tabular}
\end{tiny}
\end{table}


\bsp	
\label{lastpage}
\end{document}